\documentclass[prx,reprint,showpacs,superscriptaddress,notitlepage,floatfix]{revtex4-2}

\usepackage{color}
\usepackage{amssymb,amsmath}
\usepackage{graphicx}
\usepackage{dcolumn}
\usepackage{multirow}
\usepackage{hyperref}
\usepackage[dvipsnames]{xcolor}
\usepackage{siunitx}	
\usepackage{soul}
\usepackage[T1]{fontenc}

\begin{document}

\title{Utilizing multimodal microscopy to reconstruct Si/SiGe interfacial atomic disorder and infer its impacts on qubit variability}









\author{Luis Fabián Peña$^*$}
\affiliation{Sandia National Laboratories, Albuquerque NM, 87185 USA}
\altaffiliation{Current affiliation: Centre of Excellence for Quantum Computation and Communication Technology, School of Physics, University of New South Wales, Sydney, NSW 2052, Australia}
\author{Justine C. Koepke}
\affiliation{Sandia National Laboratories, Albuquerque NM, 87185 USA}
\author{J. Houston Dycus}
\affiliation{Advanced Microscopy, Eurofins EAG Materials Science, Raleigh, NC USA}
\author{Andrew Mounce}
\affiliation{Center for Integrated Nanotechnologies, Sandia National Laboratories, Albuquerque NM, 87185 USA}
\author{Andrew D. Baczewski}
\affiliation{Center for Computing Research, Sandia National Laboratories, Albuquerque NM, 87185 USA}
\author{N. Tobias Jacobson}
\affiliation{Center for Computing Research, Sandia National Laboratories, Albuquerque NM, 87185 USA}
\email{ntjacob@sandia.gov}
\author{Ezra Bussmann}
\affiliation{Center for Integrated Nanotechnologies, Sandia National Laboratories, Albuquerque NM, 87185 USA}
\email{ebussma@sandia.gov}
\date{\today}



\begin{abstract}
SiGe heteroepitaxial growth yields pristine host material for quantum dot qubits, but residual interface disorder can lead to qubit-to-qubit variability that might pose an obstacle to reliable SiGe-based quantum computing. We demonstrate a technique to reconstruct 3D interfacial atomic structure spanning multiqubit areas by combining data from two verifiably atomic-resolution microscopy techniques. Utilizing scanning tunneling microscopy (STM) to track molecular beam epitaxy (MBE) growth, we image surface atomic structure following deposition of each heterostructure layer revealing nanosized SiGe undulations, disordered strained-Si atomic steps, and nonconformal uncorrelated roughness between interfaces. Since phenomena such as atomic intermixing during subsequent overgrowth inevitably modify interfaces, we measure post-growth structure via cross-sectional high-angle annular dark field scanning transmission electron microscopy (HAADF-STEM). Features such as nanosized roughness remain intact, but atomic step structure is indiscernible in $1.0\pm 0.4$~nm-wide intermixing at interfaces. Convolving STM and HAADF-STEM data yields 3D structures capturing interface roughness and intermixing. We utilize the structures in an atomistic multivalley effective mass theory to quantify qubit spectral variability. The results indicate (1) appreciable valley splitting (VS) variability of roughly $\pm$ $50\%$ owing to alloy disorder, and (2) roughness-induced double-dot detuning bias energy variability of order $1-10$ meV depending on well thickness. For measured intermixing, atomic steps have negligible influence on VS, and uncorrelated roughness causes spatially fluctuating energy biases in double-dot detunings potentially incorrectly attributed to charge disorder. Our approach yields atomic structure spanning orders-of-magnitude larger areas than post-growth microscopy or tomography alone enabling more holistic predictions of disorder-induced qubit variability.
\end{abstract}

\maketitle{}
\section{Introduction}
\begin{figure}
\includegraphics[width=3.25in]{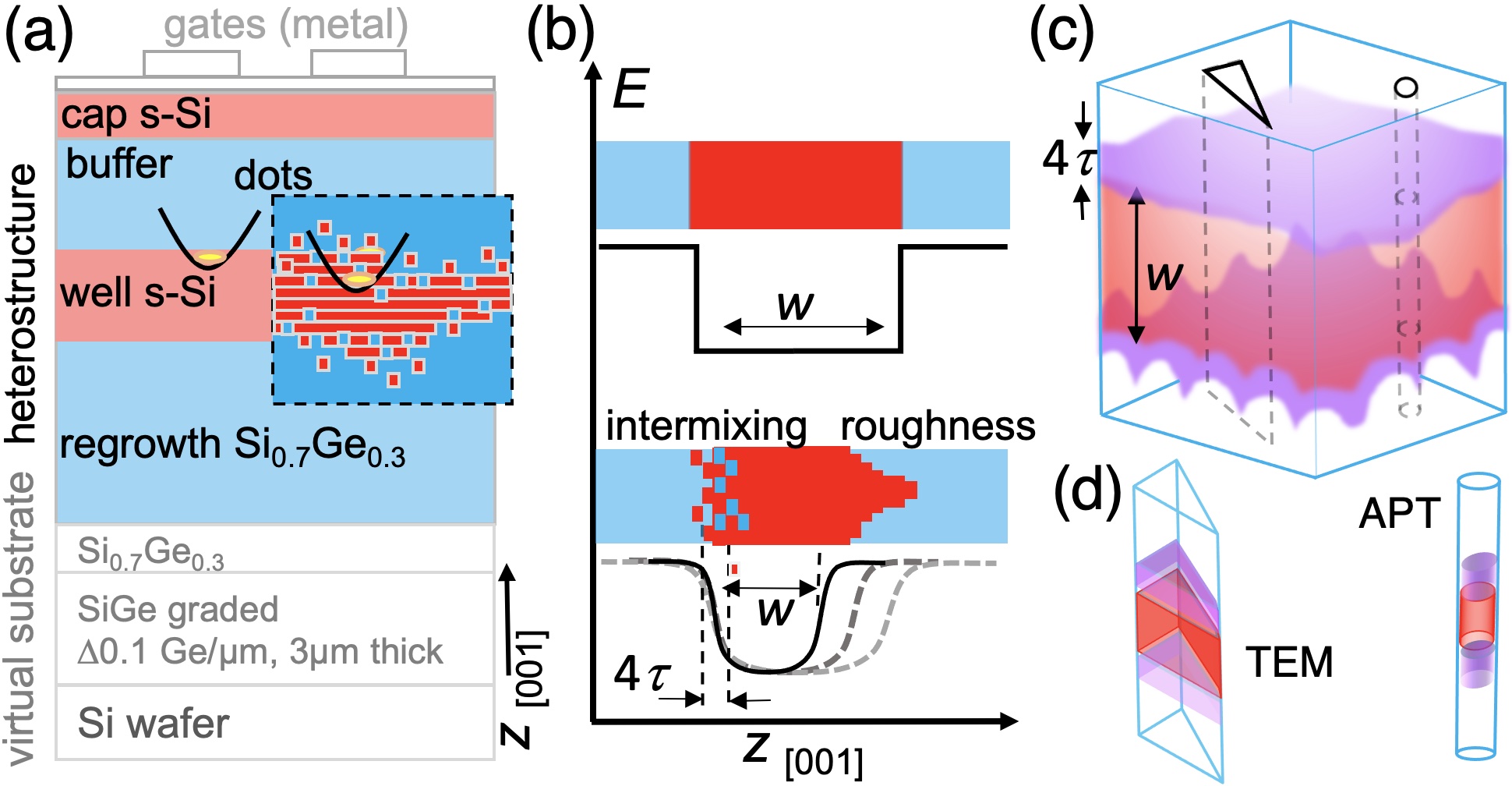}
\caption{(a) A schematic of the heterostructure in this work consisting of a s-Si well layer embedded within SiGe layers.  The inset (within the dashed rectangle) depicts intermixing and roughness disorder near interfaces. (b) An energy $E$ diagram indicating that a quantum well forms in the s-Si layer owing to conduction band edge offset from adjacent SiGe layers, and the confinement potential width, $w$, and transition length ($\sim4\tau$) are shaped by growth roughness and intermixing. Generally, interface roughness is not strictly correlated between interfaces and intermixing may be superposed to yield (c) complex 3D interface structures that are intractable for (d) the highest spatial resolution post-growth measurement techniques, such as transmission electron microscopy (TEM) and APT owing to, e.g., limited practical measurement volumes smaller than established roughness autocorrelation lengths, along with probe convolution effects, e.g., averaging of structure information along the TEM beam path through a cross-sectional slice.}
\label{fig1} 
\end{figure}

Nanoelectronic devices using Si/SiGe heterostructures to host quantum dot qubits offer robust coherence, one-/two-qubit gate fidelity, and compact device footprints compatible with Si foundry processing.\cite{maune2012,Zwanenburg2013,Kawakami2014,zajac2016,yoneda2018,yang2019,mills2022,noiri2022, Xue2022,Philips2022,Takeda2022,Weinstein2023} With the ultimate goal of monolithic Si integration, recent qubit research primarily utilizes epitaxial single quantum well heterostructures depicted schematically in Fig.~\ref{fig1}~(a).\cite{Sakr2005,Goswami2007,Borselli2011,lu2011,Schaffler1997, richardson2016,scappucci2021,gyure2021} Briefly, typical qubit heterostructure material comprises a strained-Si (s-Si) well layer pseudomorphically lattice-matched in-plane with surrounding relaxed Si$_{1-x}$Ge$_{x}$, $x\sim0.3$.\cite{richardson2016,scappucci2021} Leading qubit varieties consist of two or three coupled electrostatic dots, depicted as harmonic wells in Fig.~\ref{fig1}~(a), confining one or a few electrons vertically in the s-Si well by type-II band offsets, Fig.~\ref{fig1}~(b), and laterally by voltages on nanoscale metal gates [top Fig.~\ref{fig1}~(a)].\cite{Eng2015,gyure2021} Heterostructure growth by chemical vapor deposition (CVD) and molecular beam epitaxy (MBE) yields suitable qubit environments in the s-Si well with figures-of-merit including low metal-insulator percolation e$^{-}$ densities (<$10^{11}$ cm$^{-2}$), and minimal nuclear spin background via $^{28}$Si (spin-free) isotopic enrichment (>$99.9\%$).\cite{Schaffler1997,richardson2016,scappucci2021,gyure2021} Consequently, this material has enabled leading Si-based qubit technology demonstrations, e.g., coupling multiple high-fidelity qubits and rudimentary quantum error correction.\cite{mills2022,noiri2022, Xue2022,Philips2022,Takeda2022,Weinstein2023} Investigation and understanding of salient future scale-up challenges including expected variability over qubit ensembles is timely.\cite{Dodson2022, McJunkin2022}

Residual Si/SiGe interfacial atomic structure disorder is one cause for qubit variability.\cite{Neyens2018,Tariq2019,McJunkin2021b,PaqueletWuetz2022} In contrast to the ideal of flat interfaces and abrupt potentials, realistic structure includes disorder, inset right side Fig.~\ref{fig1}~(a), and resulting variability in qubit confinement potentials, bottom Fig.~\ref{fig1}~(b).\cite{Zwanenburg2013} Disorder-induced qubit variability might result from intimate contact between dot electron wave functions and disordered interfaces exhibiting (1) random intermixing between miscible Si and Ge and (2) growth roughness of interfaces between the Si and SiGe layers. Factor (1) results in broader interface barrier potential, bottom Fig.~\ref{fig1}~(b), and consequent dot-to-dot variability in the valley character of low-lying orbitals, impacting both valley splitting (VS) and inter-dot tunnel or exchange couplings used to drive logic gating operations.\cite{Mi2017,Neyens2018,Borjans2021} Factor (2) modifies the quantum well width, $w$, in Fig.~\ref{fig1}~(b), resulting in appreciable variation of inter-dot energy biases that might otherwise be attributed to charge disorder. For qubit operation protocols that may depend on tight control of quantum dot energy offsets, such as ``conveyor mode'' shuttling, this might serve as a nuisance source of variation that is similar in effect to but distinct in origin from charge disorder.\cite{Seidler2022,Burkard2021}
 
Valley splitting is a focal point for experiments, theory, and simulations because experiments indicate VS variability in the range $20-300$~$\mu$eV.\cite{Borselli2011,Zajac2015,Scarlino2017,Mi2018,Borjans2019,Hollmann2020,Chen2021,PaqueletWuetz2022,Philips2022} This variability is significant because it is comparable to the energies separating typical qubit basis states. So, valley states are a degree-of-freedom that can act as a potential leakage channel or be harnessed into new forms of qubits.\cite{Borselli2011,Zajac2015,Scarlino2017,Mi2018,Borjans2019,Hollmann2020,Bussmann2018,Chen2021,PaqueletWuetz2022} In either case, it is useful to understand and quantify VS variability.

To understand and discover control strategies for qubit-to-qubit VS variability, empirical pseudopotential, tight-binding, and effective mass calculations have been useful tools.\cite{Zwanenburg2013,Burkard2021} Early work assessed VS variability starting from principled assumptions that as yet unresolved embedded interface structure consists of features, e.g., discrete mono-/bilayer- atomic-step roughness, seen on s-Si and SiGe growth surfaces with local miscuts, Fig.~\ref{fig1}~(c).\cite{Boykin2004,Boykin2004c,Kharche2007,Friesen2010,Zhang2013,Zwanenburg2013} More recently, near-atomic-resolution studies using atom-probe tomography (APT) supplemented with high-angle annular dark field scanning transmission electron microscopy (HAADF-STEM) imaging over the 10-nm-scale, show gradual interface transitions, Fig.~\ref{fig1}~(b-c), and diffuse alloy disorder essentially ruling out abrupt stepped interfaces. Reported interface widths span 0.7-1.0$\pm$0.3 nm, i.e. several atomic layers, with alloy number fluctuations adding variability.\cite{Dyck2017,Koelling2022,PaqueletWuetz2022} Hence, recent theory, simulation, and experiment focus primarily on VS variability owing to random alloy disorder and alloy fluctuations.\cite{Neyens2018,McJunkin2022,PaqueletWuetz2022,Losert2023} 

In contrast to VS variability, which is a consequence of the particular atomic-scale alloy disorder realized in the vicinity of any given quantum dot, orbital level variability in response to disorder is comparatively straightforward to understand as a consequence of varying well width, $w$, Fig.~\ref{fig1}~(b).\cite{Zwanenburg2013} Well width is expected to vary owing to undulations, e.g., local growth roughness, at each interface. Prior work with hard X-ray nanospot diffraction shows roughly periodic (few-hundred-nm wavelength) lateral undulations of well width at a few atomic layer amplitude.\cite{Evans2012} Such long-period undulation is unlikely to be connected with (\AA-scale) alloy disorder and was attributed to epitaxial growth roughness, Fig.~\ref{fig1}~(b). Overall, the available data hints at a qualitative structural description of the s-Si well and interfaces including longer-period undulations (roughness) convolved with diffuse interface broadening (intermixing), as depicted in the perspective view in Fig.~\ref{fig1}~(c). 

Anticipating appreciable qubit-to-qubit variability owing to contributions of roughness and alloy disorder, a strategy embraced in recent works is to engineer ensemble distributions, e.g., VS distributions, over many qubits by targeted manipulation of ensemble disorder.\cite{Neyens2018,McJunkin2022,PaqueletWuetz2022,Losert2023} For example, precision placement of inherently-disordered layers of Ge in or near the well is found to amplify VS.\cite{McJunkin2022,PaqueletWuetz2022} To advance this strategy, some recent theory implementations (atomistic tight binding, empirical pseudopotential, effective mass theory) have been developed to predict structure-VS relationships from atomistic materials descriptions capturing specific disorder realizations.\cite{McJunkin2022,PaqueletWuetz2022,Wang2022} The calculations demand accurate ensemble descriptions of buried interfaces over volumes encountered by numerous qubits sampling multiple forms of disorder. Comprehensive 3D multiscale ensemble descriptions capturing both longer-ranged undulations convolved with interfacial alloy intermixing, Fig.~\ref{fig1}~(c), are intractable for individual local probing methods, e.g., HAADF-STEM and APT, Fig.~\ref{fig1}~(d), owing to limited sample volumes and image convolution effects, indicating that a combination of techniques sampling across atomic-to-micrometer interface disorder realizations will yield more complete structural descriptions.

In this manuscript, we describe a multimodal microscopy approach to reconstruct 3D atomic structure at Si/SiGe heterointerfaces. Then we use the structures to model dot-to-dot (qubit) variability of VS and orbital  levels (detunings). The results characterize interface undulation, alloy disorder, and resulting variability of dot spectral properties over areas (>$1 \mu$m$^2$) characteristic of multi-qubit devices. Utilizing scanning tunneling microscopy (STM) to track molecular beam epitaxy (MBE) growth, we image surface atomic structure over micron-square areas following deposition of each heterostructure layer. STM indicates \AA-to-nanometer roughness of s-Si and relaxed Si$_{0.7}$Ge$_{0.3}$ growth surfaces that subsequently become buried interfaces. The finished heterostructure interfaces are imaged using cross-sectional HAADF-STEM. For s-Si, STM shows atomically flat growth surfaces with Poisson-random terrace sizes cascading along the (local) miscut.\cite{Voigtlander2001} By contrast, 2D adatom island nucleation, stacking, and resulting nanometer-sized undulations dominate SiGe surfaces. Si-Ge miscibility and established intermixing pathways render uncertain what surface structures survive burial during overgrowth ($T=550^{\circ}$C). Post-growth HAADF-STEM imaging near regions probed by STM reveals that nanometer-sized SiGe roughness survives largely intact, while \AA-sized s-Si step-terrace structure is lost in diffuse $1.0\pm0.4$~nm-wide interfaces. Interfaces in STM and HAADF-STEM are reconciled assuming established intermixing mechanisms, supporting an overall structure description intractable to prior reported approaches using individual local-probe techniques, e.g., HAADF-STEM or APT alone. Notably, our STM and HAADF-STEM data reveal interface roughness autocorrelation lengths ($40-107$ nm) that are longer than the entire span ($\sim 30\times30$) of post-growth microscopy or tomography structural data used in related prior works.\cite{Dyck2017, PaqueletWuetz2022} We use our structure data to calculate dot-to-dot variability including well thickness variation-induced inter-dot energy biases and valley splitting statistics. We predict a spread of electronic conduction band valley splittings of $0-200\mu$eV, which is consistent with available experimental VS measurements on qubits.\cite{Borselli2011, Zajac2015, Scarlino2017, Mi2018, Borjans2019, Hollmann2020, Chen2021, PaqueletWuetz2022,Philips2022} Also, we use our data to estimate inter-dot bias spatial variability owing to interface roughness modulating quantum well confinement along the growth axis by up to an appreciable tens of meV for typical well thicknesses ($5-10$ nm). We are not aware of such a broad-range ensemble structure-properties description at the atomic resolution limit in both measurement and modeling having been reported previously. Both model findings related to VS and orbital variability have significance for understanding performance limits in quantum computing applications.

\section{Experiment, SiGe growth \& atomic structure measurements}
\begin{figure*}
 \includegraphics[width=5.5in]{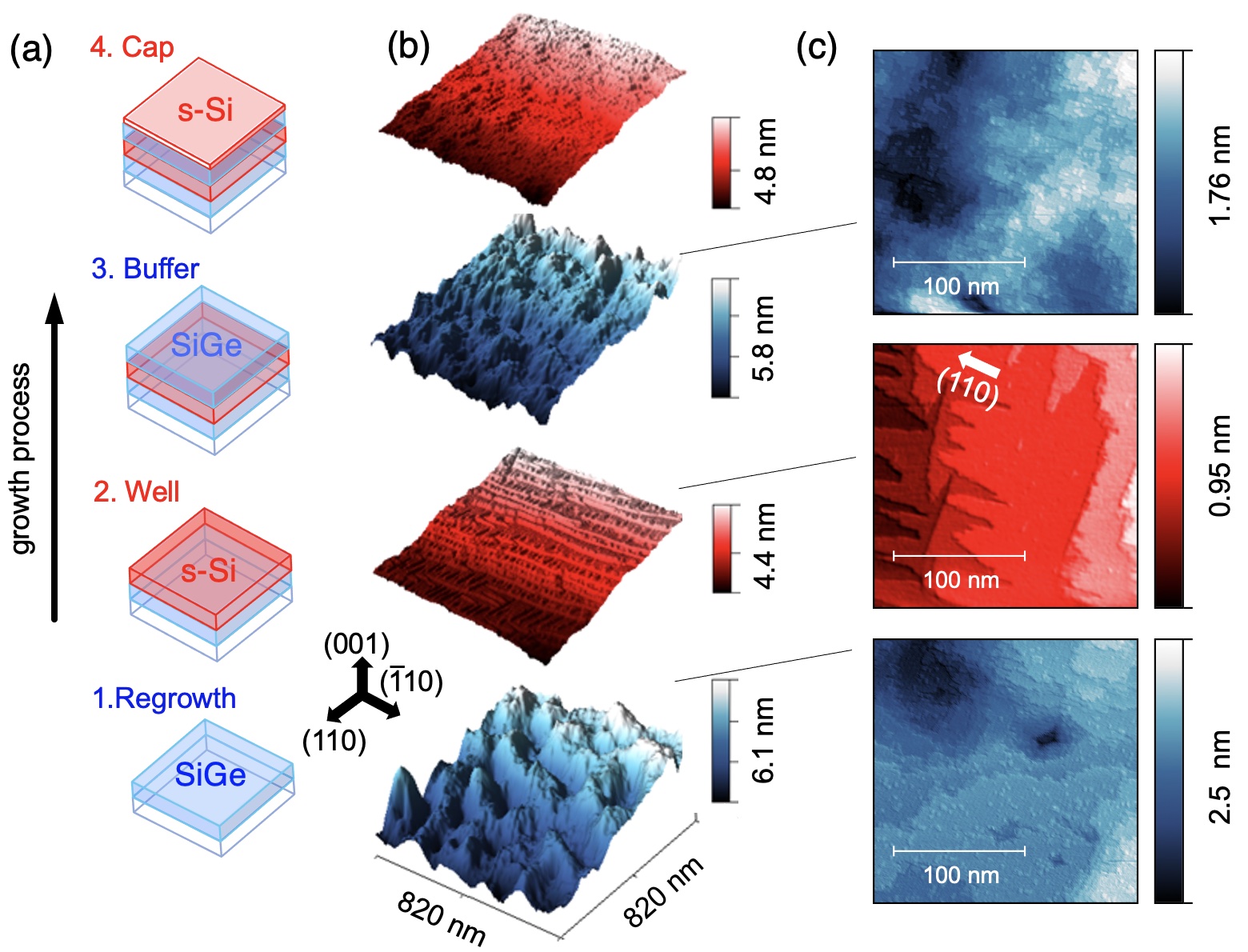}
\caption{The heterostructure layer sequence. Each column of this figure is to be read from bottom to top and shows: (a) a schematic indicating heterostructure layer sequence, (b) STM images showing surface structure upon layer completion just prior to overgrowth and heterointerface formation, (c) STM images in smaller areas showing some key atomic structure features. Note that the area of the images in column (c) is orders of magnitude greater than the interface area typically sampled by techniques such as APT and HAADF-STEM. STM images were acquired at 3V/0.5 nA tunnel current. The data has been plotted in a coordinate system defined by principal crystal directions, as indicated, to enable accurate, oriented, geometric comparisons to HAADF-STEM data.}
\label{fig2} 
\end{figure*}

Our STM growth study follows a layer sequence in Fig.~\ref{fig2}~(a). Supplementary Information (SI) S1 describes material preparations, the growth process, and STM data acquisition and analysis (see Figs.~S1.1-1.4). Si/SiGe heterostructure MBE on relaxed, epi-ready, Si$_{0.7}$Ge$_{0.3}$ virtual substrates started by depositing a $70$~nm-thick SiGe regrowth layer surface shown in STM images in Fig.~\ref{fig2}~(b-c). The regrowth surface undulates at the nanometer scale (RMS roughness $\sim 0.5$~nm) and sparse metastable artifacts resembling pits appear occasionally, Fig.~\ref{fig2}~(c). Similar pits are associated with threading dislocations terminating at the surface, although we see no indication of dislocations reaching the heterostructure layers in post-growth HAADF-STEM images.\cite{Hsu1992,Jesson1997,Fitzgerald1997}

Next, a $\sim$15 nm thick s-Si well was deposited on the SiGe regrowth layer. Comparing STM data, Fig.~\ref{fig2}~(b-c), the well's dominant qualitative surface features are totally different from the regrowth surface. Moreover, it is evident that the layout of atomic steps at the s-Si well surface differs from the uniform staircase-like distribution commonly utilized in prior models.\cite{Zwanenburg2013}. The well surface is dominated by atomic steps with extreme fluctuations, which is an indicator of competition between underlying strain fields, step energies, and surface stress that manifests step oscillations, and even bunching in equilibrium, which enhances roughness.\cite{Tersoff1995, Jones1995, Jesson1997, Ebner1997,Alerhand1988,Tromp1992,Ebner1997} As the average terrace width decreases, the likelihood of step bunching increases.\cite{Swartzentruber1993} 

A smoothing effect was observed after Si epitaxy, consistent with its inherent step-flow growth mechanism at $T\ge 550^{\circ}$C that reduces surface roughness (similarly observed by Baribeau and Kuan et al.).\cite{Kuan1991,Baribeau1995} Contrary to historical model assumptions, for low miscut samples (miscut$\sim0.5^{\circ}$), we observed no correlation between atomic-scale surface structure from the underlying substrate and the top of the quantum well, Fig.~\ref{fig2}~(c).\cite{Zwanenburg2013} This indicates that initial surface topography does not translate to subsequent layers, and new material does not grow conformally but instead is governed by each layer’s differing growth dynamics. For example, pit-like artifacts observed on the regrowth layer did not translate to the s-Si well, indicating that they are metastable in SiGe layers. These observations are captured quantitatively in a structure analysis via $z$ (height) autocorrelation functions (see SI1 Fig. S1.5 and S1.6). Autocorrelation functions indicate significantly different surface character for the SiGe regrowth versus the s-Si well surfaces, both in terms of root mean square (RMS) roughness (0.54 versus 0.18 nm, respectively) and correlation lengths (45 nm versus 107 nm, respectively), indicating no evident relationship of surface morphology. The apparent lack of correlation from interface to interface is most qualitatively consistent with Evans et al.'s X-ray data.\cite{Evans2012} Notably, uncorrelated interface behavior indicates that well thickness is likely to vary spatially, leading to variability of qubit (dot) orbital levels.\cite{Evans2012}
 
Following well growth, a $45$~nm-thick SiGe buffer layer was deposited. Buffer surface structure, Fig.~\ref{fig2}~(c), resembles the SiGe regrowth layer, {\it e.g.,} there are surface undulations and some sparse (metastable) pit-like artifacts.\cite{Yitamben2017} An increase in surface roughness compared to the well is evident (and summarized in SI1 Fig. S1.6).\cite{Qin2000, Schelling2000}
Finally, a 3-nm-thick Si cap was deposited as a seed layer for subsequent atomic layer deposition of a dielectric for metal gate isolation. It is important to point out that this final surface is not evidently correlated with the surface structure of the s-Si well, i.e., the random atomic step order differs, Fig.~\ref{fig2}~(b). Thus, post-growth surface measurements, e.g. atomic force microscopy, of the final surface will not correlate very closely with surface structure at the buried s-Si well or other layers.

So far, we have described growth surfaces observed via STM after each new heterostructure layer is added, and just prior to overgrowth and heterointerface formation. Since surface dynamics including step mobility and intermixing are integral to growth, it is nearly certain that surfaces that we observed with STM change through interface formation.\cite{Qin2000,Uberuaga2000,Hannon2004, Bussmann2010}

\begin{figure*}
 \includegraphics[width=5.5in]{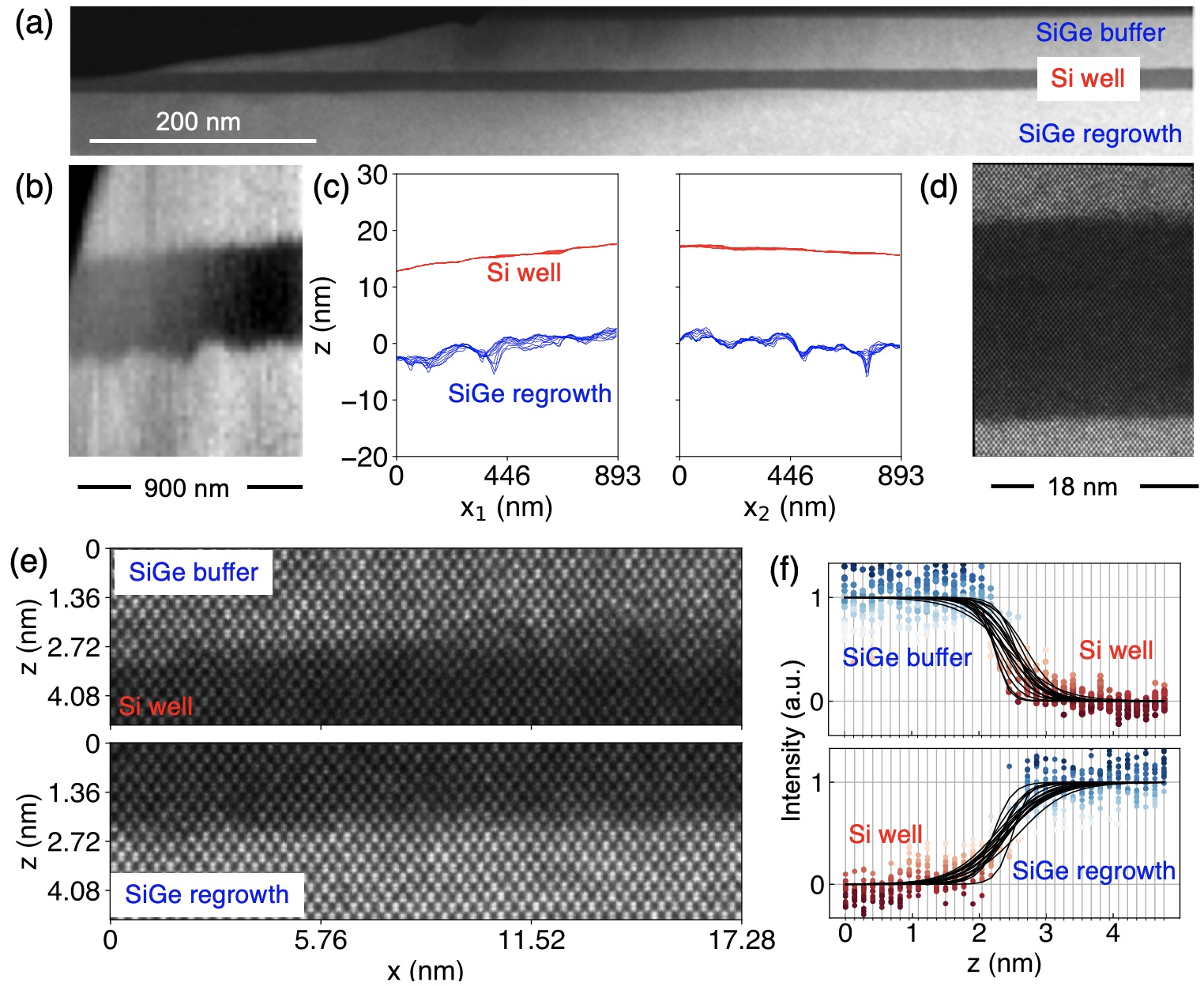}
\caption{(a) Post-growth cross-sectional HAADF-STEM image of the heterostructure. Note: STEM-EELS indicates cross-section thickness (normal to page) ranges from 20-50 nm thick from left-to-right. (b) HAADF data plotted with $\sim 10:1$ vertical:horizontal scaling highlights nanosize structure similar to (c) STM data plotted on the same scale, in the same crystal-oriented coordinate system, along  $x_{i=1,2}=$~[110]-equivalent directions. Comparing HAADF-STEM images with STM data shows nanosized growth surface and buried interface features are very similar. (d) Atomic-resolution HAADF-STEM of the s-Si well, (e) cropped to show heterointerfacial atomic details indicating that the interfaces are gradual and intermixed at the atomic scale. Here, STEM-EELS (SI2) indicates 22 nm lamella thickness. (f) Plots of column intensity across the interface span approximately $7\pm3$ atomic layers on average across the image, after correcting for HAADF-STEM beam spread within the solid. The method of interface width measurement is indicated in SI2.}
\label{fig3} 
\end{figure*}

To compare STM surface structure with buried interface structure, we utilize post-growth cross-sectional HAADF-STEM scanning transmission electron microscopy along [110] directions. Details of HAADF-STEM lamella preparation, measurement, and analysis are described in Supplementary Information (SI) S2. The cross-sectional lamella imaged here is cut from an interface region that is near (within same square millimeter) but not identical to STM data in Fig.~\ref{fig2}~(b). The lamella is wedge-shaped tapering from roughly $20-120$ nm-thick as assessed by scanning transmission electron microscopy electron energy loss spectrometry (STEM-EELS)(see SI2 Fig. S2.1). Fig.~\ref{fig3}~(a) shows a HAADF-STEM image of the s-Si well interface with adjacent SiGe layers. HAADF-STEM intensity, $I$, is a probe of nuclear charge, $Z$, with $I\sim Z^{1.8}$, so the image contrast is an indicator for Ge content in each atomic column along the electron beam path.\cite{Dyck2017} To emphasize nanosized interface features, the HAADF-STEM data is plotted in Fig.~\ref{fig3}~(b) with a $\sim$1:10 vertical stretch (see SI2 Fig. S2.2 and related discussion). Note that exaggerated vertical scaling is typical for all STM data to emphasize atomic steps and roughness, $cf$ x {\it versus} z-scales in Fig.~\ref{fig2}~(b). The HAADF-STEM images reveal: (1) that nanosized SiGe regrowth surface undulations survive overgrowth/burial, and (2) as anticipated, the s-Si well surface is relatively flat with some small undulations ($\sim$1 nm) reminiscent of waviness due to step-density fluctuations and bunches observed in STM data, Fig.~\ref{fig2}~(b-c). 

As a reference for comparing STM and HAADF-STEM image features, we have plotted several STM line traces from both the $[110]$ and $[\bar{1}10]$ directions, Fig.~\ref{fig3}~(c), encompassing roughly the same interface area ($\sim900 $~nm length $\times 47$~nm thickness, into the plane of the page) as the HAADF-STEM, and it is clear that nanoscale roughness at buried interfaces is similar to the STM growth surface roughness. Consistent with STM observations, the most significant observation from Fig.~\ref{fig3}~(a-b) is that the well thickness varies appreciably owing to roughness of the SiGe regrowth relative to the nearly atomically flat Si. 

In addition, we have calculated and compared autocorrelation functions describing the STM (growth surface) and HAADF-STEM (post-growth interface) structure (SI2 Fig.~S2.3). Consistent with the visually apparent similarity in the surface and interface data, the surface/interface autocorrelation traits (RMS roughness and correlation length) are similar, except for a significantly smaller post-growth correlation length (41 nm versus 107 nm) and an increased roughness (0.25 nm versus 0.18 nm) observed for the s-Si well top interface. We attribute this decrease in correlation length and increase in roughness to disordering effects of intermixing described in detail next.

In contrast to our STM data, atomic-resolution HAADF-STEM does not show any atomically abrupt interfaces or atomic step-terrace structure, but rather $\sim$1 nm-wide gradual diffuse interfaces in several atomic-resolution images of various lamella thicknesses. Fig.~\ref{fig3}~(d) shows one example of an atomic-resolution HAADF-STEM image, and Fig.~\ref{fig3}~(e) shows atomic detail of the gradual Si-SiGe transitions typical of our HAADF-STEM data, as well as other recent atomic-resolution HAADF-STEM and APT.\cite{Dyck2017,PaqueletWuetz2022} From our STM spatial distribution of steps (See SI1 Fig.~S1.7-1.9), we would expect to observe atomically-abrupt terraces (0.134 nm wide interfaces) over terrace regions with a finite probability of capturing at least a few steps in an area the size in Fig.~\ref{fig3}~(e). Instead, as indicated in Fig.~\ref{fig3}~(f), we observe smooth interface transitions spanning several atomic layers with smooth transitions reasonably estimated by sigmoid curves (see SI2 Figs. S2.4-2.8 for analysis details). The metric that we use to measure the interface width is $4\tau$, where $\tau$ is the sigmoid width parameter. The parameter $4\tau$ measures the distance for $\sim 0.12-0.88$ of the full transition, and is a commonly utilized and easy-to-automate tool to estimate HAADF-STEM and APT interface widths.\cite{Dyck2017,PaqueletWuetz2022}

Si and Ge intermixing between layers during overgrowth is a plausible cause for nanometer-wide interfaces in HAADF-STEM. This interpretation is consistent with established Si-Ge bulk miscibility accessed via thermally-activated surface and near-surface intermixing pathways likely to be operative at $T\leq 550^{\circ}$C.\cite{Qin2000,Uberuaga2000,Hannon2004,Bussmann2010} However, preexisting tilts and nanosized-roughness present on growth surfaces in STM, Fig.~\ref{fig2}~(b-c), and subsequently in HAADF-STEM, Fig.~\ref{fig3}~(a-b), both contribute to the apparent cross-sectional interface width since the HAADF-STEM beam interacts with and convolves embedded structure along its path through the lamella.

\begin{figure*}
 \includegraphics[width=5.5in]{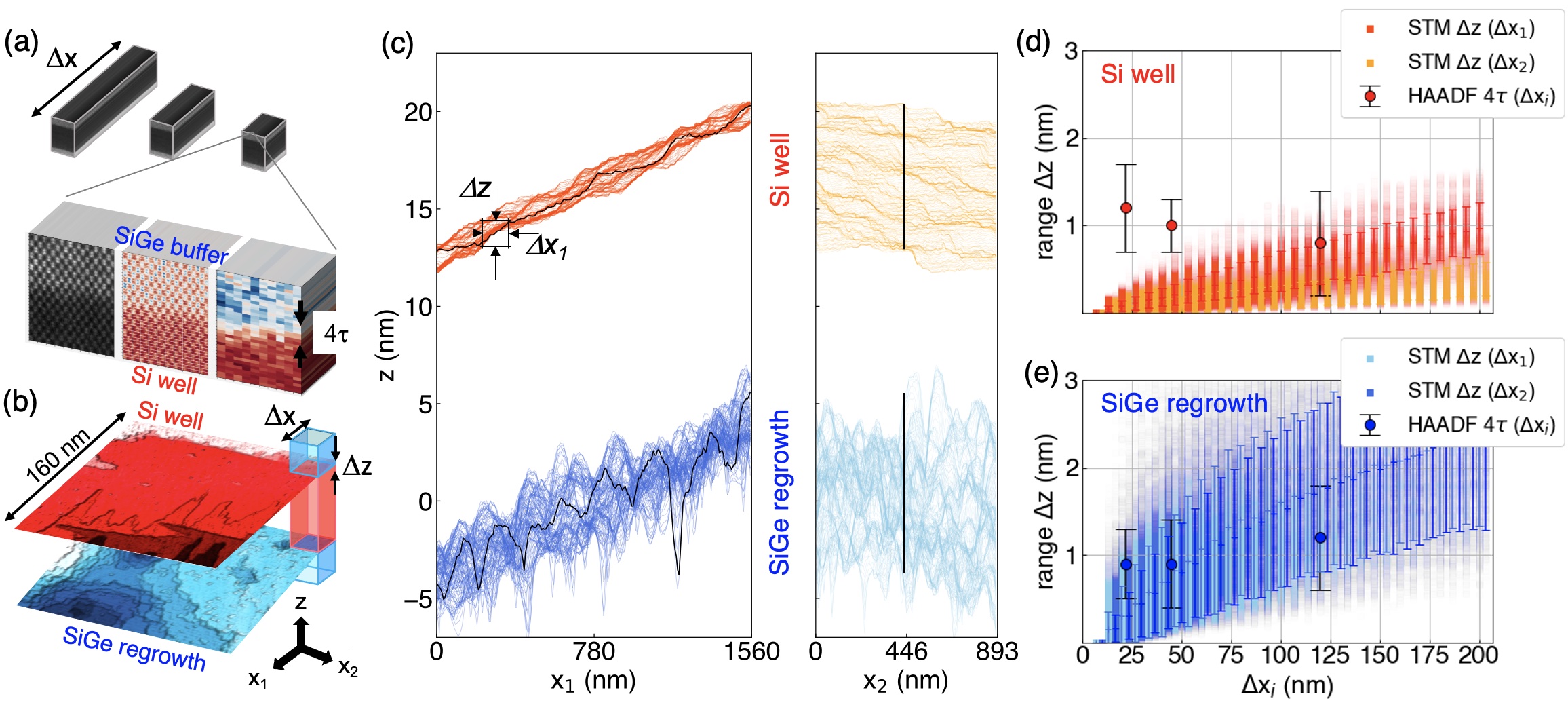}
\caption{Analysis of interface width in cross-sectional HAADF-STEM images. (a) Schematic view of our technique to measure interface width showing the approximate volumes imaged with the electron beam directed along $\Delta x$ through lamella thicknesses of 22, 45, and 120 nm. Each image is discretized by atomic columns and a measure taken of Z (atomic number) contrast, all atomic columns are fit using an automated sigmoid-fitting process, then the value $4\tau$ (see SI2) is taken as a measure of interface width. (b) Schematic showing our approach for estimating an interface-width contribution, $\Delta z$, due to roughness in STM data. (c) Two orthogonal views of the STM data used to calculate ensemble roughness, plotted in the crystal-oriented coordinates with (001) as the z-axis and (110)-directions aligned to $\Delta x_{i=1,2}$. A single line is plotted in black to indicate directional relationships in the two orthogonal views.  (d) s-Si well-to-buffer interface width vs. lamella thickness ($\Delta x$) characterized by HAADF-STEM $4\tau$ and surface width contributions, $\Delta$z, from STM data along $\Delta x_{i=1,2}$ orientations. (e) SiGe regrowth-to-Si well interface width vs. lamella thickness ($\Delta x$) characterized by HAADF-STEM 4$\tau$ with $\Delta z$ from STM data.}
\label{fig4} 
\end{figure*}

In order to discern alloy intermixing-induced broadening requires deconvolving it from other factors that potentially contribute to apparent interface width, e.g. preexisting crystal tilts and roughness seen in STM, in Fig.~\ref{fig2}~(b-c). We take an approach outlined in Fig.~\ref{fig4}~(a-b). In principle, preexisting roughness contributions can be deconvolved by performing HAADF-STEM with sufficiently thin lamella thicknesses that only the local alloy distribution contributes to the interface width. To estimate the scale for which roughness and tilt contributions to the width become negligible, we take an approach of measuring the HAADF-STEM cross-section interface width as a function of a few lamella thicknesses (22, 45, and 120 nm, see SI2 Fig. S2.4) and comparing the interface width values to contributions from preexisting tilts and roughness calculated from STM data. The approximate HAADF-STEM sample volumes are depicted in Fig.~\ref{fig4}~(a). To get an ensemble measurement of the interface width from each HAADF-STEM image area, we apply an image segmentation and sigmoid fitting routine, shown schematically in Fig.~\ref{fig4}~(a), to extract an interface width, $4\tau$, from every atomic plane in all atomic-resolution HAADF-STEM images [see SI2 Fig.~S2.8]. For conciseness, we use a mean $4\tau \pm \sigma$, where sigma denotes the standard deviation from the mean for each lamella thickness.

Next, we estimate interface width contributions due to preexisting surface roughness and tilts in STM data. We use a metric $\Delta z = \mathrm{max}(z)-\mathrm{min}(z)$, that we refer to as the range function, calculated over intervals, $\Delta x_{i=1,2}$ along $(110)$-equivalent directions in the STM data, i.e. along the same directions probed by HAADF-STEM. The concept of the STM tilt and roughness characterization for comparison to HAADF-STEM is shown in Fig.~\ref{fig4}~(b). We take $\Delta x_{i=1,2}$ to span the HAADF-STEM lamella thickness range. By rastering the $\Delta x_{i=1,2}$ window over micron-sized STM data, Fig.~\ref{fig4}~(c), a micron-scale characteristic ensemble description for $\Delta z$ is calculated along $\Delta x_{i=1,2}$ with results shown in Fig.~\ref{fig4}~(d-e). 

Finally, we place bounds on the interface intermixing length, $L$, based on our data in Fig.~\ref{fig4}. In Fig.~\ref{fig4}~(d) and (e), we see that only two HAADF-STEM interface widths from the thinnest ($22$ nm and $45$~nm-thick) lamella for the s-Si well surface are clearly distinguishable from the cloud of $\Delta z$ values. By contrast, all HAADF-STEM interface widths for the SiGe regrowth surface fall within the range of nanosized roughness and tilt observed in STM data. As we noted earlier, nanosized roughness (undulation) of SiGe surfaces is evident in post-growth HAADF-STEM data, cf. Fig.~\ref{fig3}~(a), and it is probable that we are observing these preexisting features in the HAADF-STEM for all lamella thickness. From the thinnest samples (22 and 45 nm thick) of the s-Si well interface we measure $4\tau=1.2\pm0.5$ nm and $1.0\pm0.3$ nm respectively, while growth roughness and tilt contributions are on the order of $0.04-0.08$ nm, and $0.1-0.25$ nm, respectively depending upon direction. Since the STM surfaces are atomically flat, and $\Delta z$ in STM is largely dominated by the tilt [i.e. overall slope Fig.~\ref{fig4}~(c)] $\Delta z$ increases linearly with $\Delta x_{i=1,2}$ and contributes directly to interface thickness. However, for approximate intermixing length bounds, we take the difference (in quadrature) $L = [4\tau^2-\Delta z^{2}]^{1/2}$, where the $\Delta z$ values $(0.04-0.25 ~ \mathrm{nm})$ contribute negligibly, and we conclude that $0.9$~nm~$< L < 1.2$~nm ($\pm 0.4$~nm) or $7-9 (\pm 3)$~layers.

Si and Ge are fully bulk miscible, and there are a few surface/near-surface mechanisms that are likely to be active allowing Ge to intermix with the Si over multiple atomic layers, as well as allowing Si transport upward from the original interface.\cite{Jernigan1996,Uberuaga2000,Qin2000,Boguslawski2002,Lu2002,Hannon2004,Akis2005,Zipoli2008, Bussmann2010} First, there is a near-surface enhanced interstitial mechanism of Uberuaga et al. that allows transport of appreciable Ge up to 4 atomic layers below the original Si growth surface for $T\leq500^{\circ}$C.\cite{Uberuaga2000} Second, intuitively consistent with bulk Si-Ge miscibility, SiGe alloys form a 2D surface wetting layer on Si surfaces by purely surface atomic exchange diffusion processes, leading to an upward exchange of atoms from surface lattice sites to the supersaturated gas of adatoms involved in growth, and ultimately into subsequent atomic layers as they nucleate. Surface exchange diffusion is rapid at $T>90^{\circ}$C for Si and Ge and anticipated to contribute significantly to intermixing at increasing temperatures.\cite{Qin2000,Boguslawski2002,Bussmann2010} If atomic exchange-diffusion promotes Si upward with a probability $p\sim 0.5$, then we anticipate additional Si to be distributed upward to the n$^{th}$ layer above the original growth surface with a probability $~p^{n}$, such that layers above the original Si surface become Si-rich in a diminishing, roughly geometric, progression and plausibly contribute a few additional intermixed layers to the observed 4$\tau$.\cite{Qin2000,Boguslawski2002,Bussmann2010} Hence, we conclude that the intermixing length $7-9$ atomic layers is at least possible as an outcome of prior established surface intermixing processes during growth. Finally, note that broadening effects are not limited to the upper interface, rather they are definitively resolvable and quantifiable there for thinner lamella (22, 45 nm thick). 

\section{Interface disorder model \& resulting qubit spectral variability}

Given these STM and HAADF-STEM observations, we propose a model for atomic structure for the well where the mean position $\bar{z}(x_1,x_2)$ of each interface at a given location $(x_1,x_2)$ is set by the STM data and the elemental identity at each lattice site along atomic columns across the interface is determined by drawing from the sigmoid distribution with a width set by HAADF-STEM data. In our atomistic multi-valley effective mass theory simulations, for any given alloy realization we construct a bulk silicon (diamond) lattice encompassing the simulation domain and then update the Si or Ge identity of each lattice site according to the above distribution. These structures are available from the authors on request.

\begin{figure*}
 \includegraphics[width=5.5in]{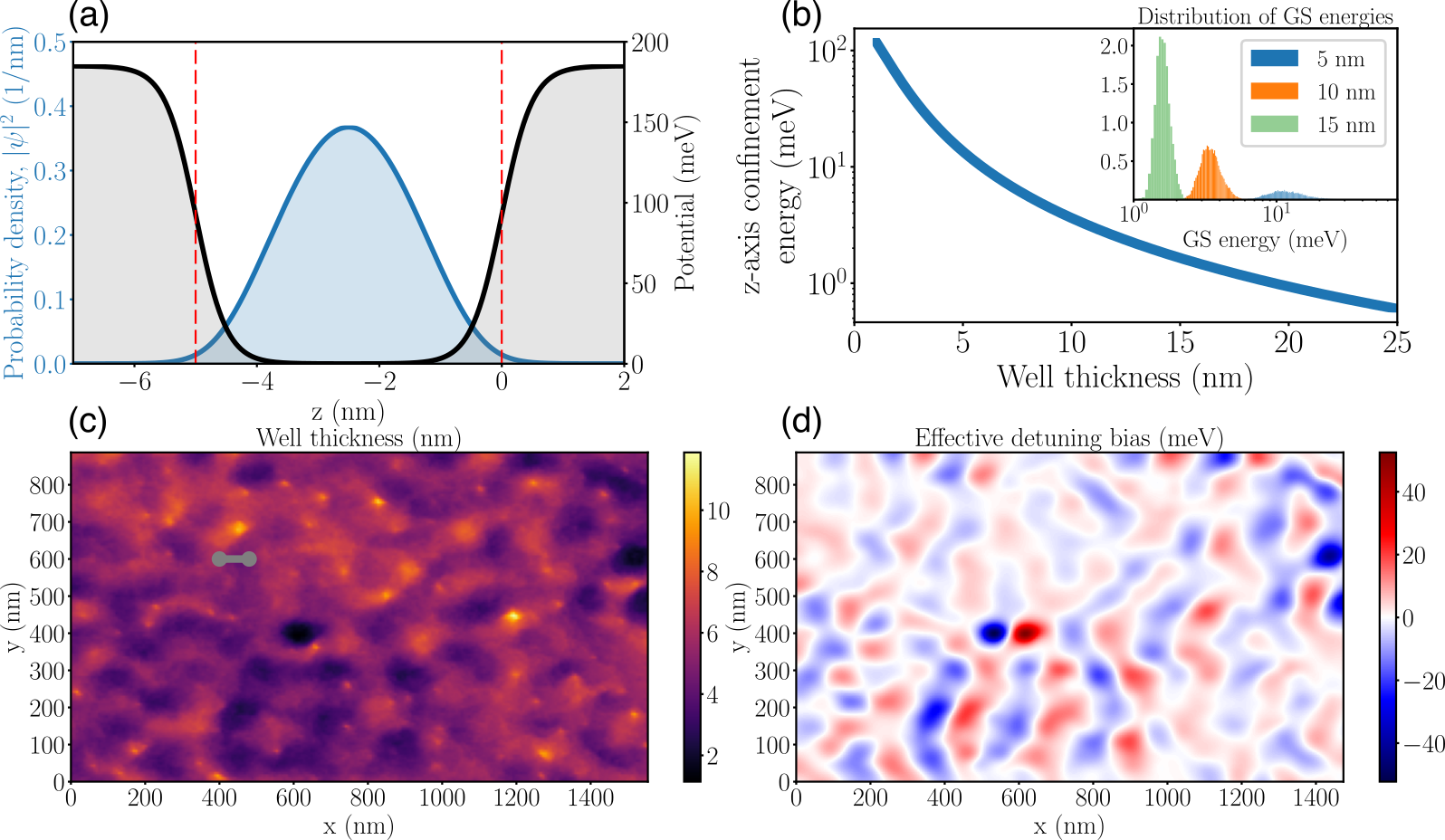}
\caption{Influence of interface topography variation and confinement along the growth axis. (a) Example one-dimensional (valley-free) Schrödinger solve for the ground state, illustrating quantum confinement along the growth axis (z-axis) of the well. Here, the quantum well is 5 nm thick, with an intermixing length of 4$\tau=$1~nm. (b) Confinement energy as a function of well thickness for 4$\tau=$ 1~nm (inset) Distribution of growth-axis confinement energies for the measured interface topography, assuming three different average well thicknesses. (c) Well thickness as a function of x,y position in the measured sample, assuming an average thickness of 5 nm and topography measured for the 15 nm sample. A pair of quantum dots 30 nm in diameter and 80 nm apart is denoted in gray to give a sense of scale. (d) Calculated effective detuning bias between dots 80 nm apart in a 5 nm well due to spatial variation of the growth-axis confinement energy.}
\label{fig5} 
\end{figure*}

Interface topographic data from both HAADF-STEM (Fig.~\ref{fig3}~(a-b)) and STM (Fig.~\ref{fig3}~(c)) indicate that the well thickness varies significantly across the sample, on the scale of a few nm. For example, the well width in the HAADF-STEM data  (Fig.~\ref{fig3}~(a-b)) fluctuates by a root mean square deviation of 1.7 nm. The well thickness sets the energy scale of confinement along the growth axis of an electron in a quantum dot. For a given quantum dot this amounts to an overall offset that, for a double- or multi-quantum dot system, will manifest as an inter-dot energy offset (detuning) bias. To model the consequences of well thickness variation, we solve the one-dimensional Schrödinger equation for various well thicknesses, including the potential induced by the conduction band offsets between the s-Si well and SiGe layers as well as interface thickness $4\tau$, assumed here to be 1 nm (Fig.~\ref{fig5}~(a)). The confinement energy as a function of well thickness is shown in Fig.~\ref{fig5}~(b). We find that the scale of variation of vertical confinement depends significantly on the mean well thickness, with thinner wells manifesting much larger fluctuations in vertical confinement. To estimate the effect of well thickness variation on the detuning bias offset between nearby double quantum dots, we use the measured variation of well thickness shown in Fig.~\ref{fig5}~(c) and assumption of a 5 nm average well thickness to find the detuning bias variation of Fig.~\ref{fig5}~(d) for quantum dots having a nominal 80 nm center-to-center separation. We account for the finite size of the dots by performing a Gaussian convolution over the x-y variation of well thickness with a standard deviation of 15 nm, a typical quantum dot size. For relatively thin wells ($\sim 5$ nm), this simulated level of detuning bias variation would correspond to significant offsets in, for example, the voltage bias on applied gate electrodes required to induce an inter-dot transition of an electron. Such variation may otherwise be attributed to charge disorder, but we point out here that well thickness variation may be another source of bias variation to consider. An open question to be addressed in future work is how thick a well must be for the well thickness variation to reach the scale of what we have observed in our 15 nm thick sample. If growth of the s-Si well is closer to conformal for thinner interfaces, we would expect that variation of well thickness may be correspondingly reduced.

\begin{figure*}
 \includegraphics[width=5.5in]{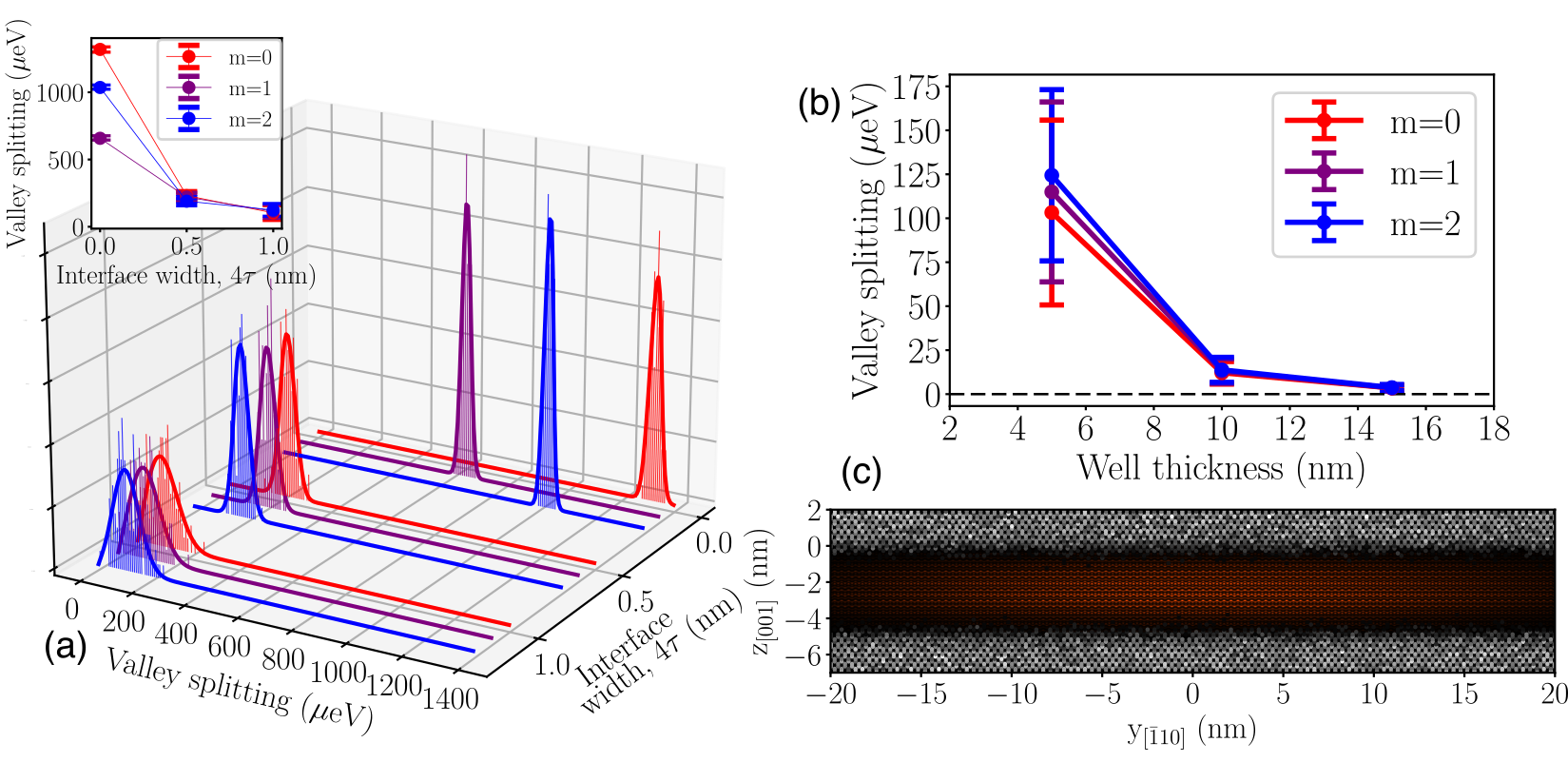}
\caption{Simulations of quantum dot valley splitting in the presence of alloy disorder and interfacial steps, assuming in-plane harmonic confinement corresponding to an orbital splitting of 1.5 meV. (a) For a 5 nm thick well, distributions of valley splitting as a function of interface width (intermixing length) 4$\tau$ in the presence of either no step (m=0), a single monoatomic layer (m=1), or two monoatomic layer (m=2) steps through the middle of the dot oriented along the [110] crystallographic axis. The solid curves are best-fit Rice distributions.\cite{PaqueletWuetz2022} (b) Valley splitting statistics as a function of well thickness for each of these step configurations, (c) Example simulation of quantum dot probability density $|\Psi|^2$ for a 5nm well, superimposed on a simulated HAADF-STEM image of Ge concentrations for a lamella 20 nm thick parallel to the [110] plane.}
\label{fig6} 
\end{figure*}

To probe disorder impacts on valley splitting, we perform multi-valley effective mass theory simulations of single-electron quantum dots in the presence of atomistic disorder corresponding to specific alloy realizations. Our simulation method incorporates detailed Bloch functions derived from density functional theory (DFT)\cite{Gamble2015} and treats each Ge atom in the simulation domain explicitly as a repulsive localized defect potential (see Methods section for more details). In Fig.~\ref{fig6}~(a), we show the distribution of valley splitting as a function of interface width $4\tau$ for a 5 nm thick well, for three different cases of step structure. In these calculations, we assume harmonic confinement in the $x$-$y$ plane corresponding to a 1.5 meV orbital splitting. We consider a step oriented along the $[110]$ axis that is $m$ atomic monolayers thick and passing through the center of the quantum dot. For the case of zero intermixing (perfectly abrupt interface), we find that the presence of the step modulates the valley splitting significantly, though the valley splitting remains relatively high ($\sim 1$ meV) on average. However, as the interface width grows the influence of the step rapidly vanishes, with even a relatively abrupt interface of  $4\tau = 0.5$ nm exhibiting negligible step-induced modulation of valley splitting.

Next, we explore how the valley splitting depends on well thickness. In Fig.~\ref{fig6}~(b), we show how the valley splitting statistics depend on well thickness in the presence of an m-monolayer atomic step through the center of the dot, in the case of an interface width of  $4\tau = 1$ nm. Well thickness clearly has a significant influence over valley splitting, with thinner wells clearly preferable to thicker wells and the presence of the few-monolayer step having minimal influence over the valley splitting distribution. In Fig.~\ref{fig6}~(c), we show the electronic wave function for a 5 nm well, along with a simulated HAADF-STEM image of Ge alloying analogous to Fig.~\ref{fig3}~(d). These simulations emphasize the critical importance of small well thickness and relatively abrupt interfaces in ensuring large valley splitting, while for realistic intermixing lengths the influence of few-monolayer steps is modest.

\section{Summary}

We have presented an analysis of interfacial atomic structure disorder for s-Si quantum wells bounded by Si$_{0.7}$Ge$_{0.3}$ layers. The technique combines two forms of atomic-resolution data: {\it in-operando/in-situ} STM interleaved with MBE growth, and post-growth HAADF-STEM. STM is applied to image surfaces of each heterostructure layer immediately prior to subsequent overgrowth (interface formation). The characteristics of s-Si surfaces include atomic steps cascading essentially monotonically along local miscuts indicative of a predominantly step-flow (Frank-van der Merwe) growth mode with little evidence for nucleation. The resulting atomic ``staircases'' with Poisson-distributed terrace widths reflect tendencies for step-bunching and atomic-step meandering attributable to elastic effects and growth step-attachment dynamics.\cite{Voigtlander2001} By contrast, SiGe has drastically different features dominated by a 2D nucleation, island-stacking, growth mode (qualitatively Stranski-Krastanov-like) leading to nanosized roughness and undulations.\cite{Jernigan2002} The s-Si versus SiGe surfaces reflect different growth dynamics and correlation between adjacent interfaces is negligible, explaining prior observations of thickness undulations in buried s-Si wells.\cite{Evans2012} Post-growth cross-sectional HAADF-STEM measurements indicate that nanoscale traits of flat s-Si versus undulating SiGe survives growth intact, but that interfaces appear broadened by $\sim$1.0$\pm0.4$~nm ($\sim$4-10 atomic layers). We rule out growth roughness as a cause for the observed breadth. Instead, interface breadth is due to intermixing reflecting Si-Ge miscibility that is accessible by surface/near-surface atomic kinetic paths.\cite{Uberuaga2000, Qin2000, Hannon2004, Bussmann2010} Consistent with STM and HAADF-STEM observations, we propose a simple overall atomic structure for the well where the mean position $\bar{z}(x_1,x_2)$ of each interface at a given location $(x_1,x_2)$ is set by the STM data and the elemental identity at each lattice site along atomic columns across the interface is set by drawing from a sigmoidal distribution. Our model extends beyond prior interface structure descriptions by more than two orders-of-magnitude in area ($<$100$\times$100 nm$^2$ to $>$1$\times1$~$\mu$m$^2$). Notably, we find surface (STM) and interface (HAADF-STEM) roughness autocorrelation lengths ($45$-$107$~nm) that are up to 3 times larger than the dimensions of data (APT and HAADF-STEM) used in other recent studies, underlining that our approach reveals broader information.\cite{Dyck2017,PaqueletWuetz2022} 

Finally, we utilize our structures to estimate growth-axis confinement energy and valley splitting variability for quantum dots in the s-Si layer. Confinement energy variability is calculated using a straightforward effective mass theory solution for the ground state confinement energies which vary in response to spatial undulations of well width resulting from uncorrelated roughness at Si/SiGe surfaces. We find that this interfacial roughness leads to appreciable confinement energy variability in our simulations of tens of meV for a 5-nm-thick well. This implies that each dot must be uniquely tuned adding complexity to e.g. shuttling, requiring coordinated manipulation of the electrostatic potential landscape. Valley splitting varies appreciable, e.g. in the range of 0-200 $\mu$eV for a 5-nm-thick well with $4\tau=1$~nm interface width. This significant VS variability presents similar measurement challenges. Mitigation strategies for VS variability have been proposed, e.g. positioning Ge layers strategically in the well or near interfaces to regularize valley-state phases and break degeneracies.\cite{Zwanenburg2013,McJunkin2022,PaqueletWuetz2022} We anticipate our structure data and model to be useful in understanding outcomes of such strategies.

\section{Acknowledgement}
We thank R. E. Butera and C. J. K. Richardson of University of Maryland, C. Barry Carter, Emeritus Professor U. of Connecticut, and Tzu-Ming Lu at Sandia National Laboratories for thought-provoking discussions. This work was performed at the Center for Integrated Nanotechnologies, an Office of Science User Facility operated for the U.S. Department of Energy (DOE) Office of Science. 
Research supported as part of $\mu$-ATOMS, an Energy Frontier Research Center funded by the U.S. Department of Energy (DOE), Office of Science, Basic Energy Sciences (BES), under award DE-SC0023412 (data analysis and manuscript preparation). Sandia National Laboratories is a multi-mission laboratory managed and operated by National Technology and Engineering Solutions of Sandia, LLC, a wholly owned subsidiary of Honeywell International, Inc., for the U.S. DOE’s National Nuclear Security Administration under contract DE-NA-0003525. This paper describes objective technical results and analysis. Any subjective views or opinions that might be expressed in the paper do not necessarily represent the views of the U.S. Department of Energy or the United States Government.

\section{Author Contributions}
L. F. Peña: investigation, formal analysis, writing - review and editing;
J. C. Koepke: investigation, writing - review and editing; 
A Mounce: software; 
A. D. Baczewski: software, writing - review and editing; 
N.T. Jacobson: theoretical analysis, software, writing - original draft, funding acquisition, project administration; 
E. Bussmann: conceptualization, writing - original draft, funding acquisition, project administration.

\bibliography{bibliography.bib}
\bibliographystyle{unsrt}

\end{document}